\documentclass[twocolumn,aps,pra,groupedaddress,amssymb]{revtex4}

\usepackage{natbib}
\usepackage{graphicx}
\usepackage{units}
\usepackage{braket}
\usepackage{amsmath}
\usepackage{color}
\usepackage{epstopdf}

\begin{document}
\hyphenation{Ryd-berg}

\title{Observation of mixed singlet-triplet Rb$_2$ Rydberg~molecules}

\author{F. B\"{o}ttcher}
\author{A. Gaj}
\author{K. M. Westphal}
\author{M.~Schlagm\"{u}ller}
\author{K.~S.~Kleinbach}
\author{R. L\"{o}w}
\author{T. Cubel Liebisch}
\author{T. Pfau}
\author{S. Hofferberth}
\email[Electronic address: \\]{s.hofferberth@physik.uni-stuttgart.de}
\affiliation{5. Physikalisches Institut and Center for Integrated Quantum Science
and Technology, Universit\"{a}t Stuttgart, Pfaffenwaldring 57, 70569 Stuttgart,
Germany}


\begin{abstract}
We present high-resolution spectroscopy of Rb$_\text{2}$ ultralong-range Rydberg molecules bound by mixed singlet-triplet electron-neutral atom scattering. The mixing of the scattering channels is a consequence of the hyperfine interaction in the ground-state atom, as predicted recently by Anderson et al. \cite{Anderson2014b}. Our experimental data enables the determination of the effective zero-energy singlet $s$-wave scattering length for Rb. We show that an external magnetic field can tune the contributions of the singlet and the triplet scattering channels and therefore the binding energies of the observed molecules. This mixing of molecular states via the magnetic field results in observed shifts of the molecular line which differ from the Zeeman shift of the asymptotic atomic states. Finally, we calculate molecular potentials using a full diagonalization approach including the $p$-wave contribution and all orders in the relative momentum $k$, and compare the obtained molecular binding energies to the experimental data.
\end{abstract}
\maketitle

\section{Introduction}
Scattering experiments can provide detailed information about the structure of matter and the character of interactions and have become a basic tool in nearly all branches of
fundamental physics. The scattering between an electron and a neutral atom depends on the kinetic energy of the particles \cite{Ramsauer1921} and the relative spin
orientation~\cite{Bahrim2001,Fabrikant1986}.
\\ \indent
The interaction between the valence electron of a Rydberg atom and a ground state atom can be treated as such a scattering problem \cite{Fermi1934} since the Rydberg electron is very
loosely bound to the core. The arising scattering potential can support bound states and ipso facto, lead to the formation of ultralong-range Rydberg molecules~\cite{Greene2000}.
These molecules were first observed in Rb for $S$-states with the Rydberg electron, and ground state atom spins in a triplet configuration \cite{Bendkowsky2009}. Since then a
variety of exciting phenomena have been explored, such as states bound by quantum reflection \cite{Bendkowsky2010}, coherent creation and breaking of the molecular bond \cite{Butscher2011}, polyatomic Rydberg molecules \cite{Gaj2014}, exotic trilobite states \cite{Li2011,Booth2015} and controlled hybridization of the molecular bond \cite{Gaj2015}. Diatomic Rydberg molecules were realized for $S$-states in Cs \cite{Tallant2012} and Sr \cite{DeSalvo2015}, for $D$-states in Rb \cite{Anderson2014,Krupp2014} and for $P$-states in Rb \cite{Bellos2013} and Cs \cite{Sassmannhausen2015}. Furthermore, Rb$_\text{2}$ Rydberg molecules were also used as a  probe of the quantum phase transition from the superfluid to the Mott-insulator phase \cite{Manthey2015}.
\\ \indent
Recently, the existence of Rydberg molecules bound by mixed singlet-triplet scattering has been proposed by Anderson et al.~\cite{Anderson2014b} and proven experimentally for Cs in
zero magnetic field \cite{Sassmannhausen2015}. In this paper, we present high resolution spectroscopic data of these mixed singlet-triplet molecules, photoassociated from a
sub-$\mu\text{K}$ cloud of $^{87}$Rb. The mixing of the singlet and triplet scattering channels is a consequence of the hyperfine interaction of the ground state atom as well as the applied magnetic field. Our data enables the determination of an effective singlet $s$-wave scattering length for $^{87}$Rb from fitting the resulting molecular binding energies. We also compare our experimental results to a full diagonalization of the scattering Hamiltonian using the non-relativistic $e^-$\,-\,Rb scattering phase shifts from \cite{Fabrikant1986}, providing a high-precision test of the low-energy range of these phase shifts. We find that in particular the triplet $p$-wave contribution significantly affects the binding potentials due to a shape resonance in the electron-neutral collision \cite{Bendkowsky2010}. Finally, we investigate the magnetic field-induced shift of the mixed-spin molecular states, which we find to be different than that of the asymptotic atomic states. We show that a transition from the Zeeman regime to a Paschen-Back like regime, where the magnetic field induced energy shifts are large compared to the molecular binding energies, happens already at very small fields of order $1$\,G for principal quantum numbers $n\approx 40$. Thus, perturbative treatment of the magnetic field is only applicable for significantly smaller fields \cite{Anderson2014b}.

\section{Mixed Singlet-Triplet Rydberg Molecules Theory}
In this section we present the theory of Rydberg molecules including the ground state hyperfine interaction, which leads to the mixing of singlet and triplet scattering channels \cite{Anderson2014b}. We first introduce the Hamiltonian of the Rydberg-neutral atom system and then explain the two approaches we use to calculate the molecular potentials, namely the effective zero-energy s-wave scattering length approach and an approach using the full diagonalization of the scattering Hamiltonian. Finally, we present the obtained potential energy curves (PECs) and show the importance of the p-wave shape resonance for the mixed singlet-triplet PECs.

\subsection{Rydberg electron Hamiltonian}
\begin{figure}[t]
\centering
\includegraphics[width=\columnwidth]{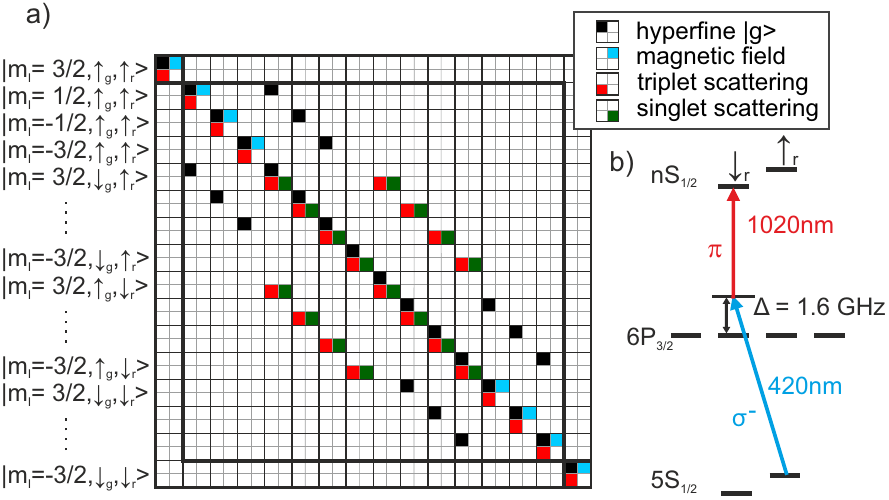}
\caption{\label{fig1} a) Schematic illustration of the matrix Hamiltonian (Eq.\,\ref{eq:1}),
where for better readability the Rydberg hyperfine interaction for a single Rydberg
$nS_{1/2}$-state is omitted. The respective terms are marked in different colors. Blank
square indicates zero in the Hamiltonian. Basis states are listed according to the tensor
operation $\hat{S}_r \otimes \hat{S}_g \otimes \hat{I}$. b) Excitation scheme to the Rydberg
state including the fine structure. The small black arrows at the top indicate the Rydberg
electron spin. $\sigma^-$ and $\pi$ denote the laser polarizations. The large detuning from
the intermediate state allows for the usage of the $j$-basis for the Rydberg state.}
\end{figure}

Within the Born-Oppenheimer approximation, we describe the interaction of the Rydberg electron with the ground state atom at position $R$ in an external magnetic field $B$ by
using the Hamiltonian:
\begin{equation}
\begin{array}{rl}
\hat{H} &= \hat{H}_0  + \hat{H}_\text{B}  +   \hat{H}_\text{HF,g} +   \hat{H}_\text{HF,r}  + V_\text{T}(R) \cdot \hat{P}_\text{T} \\ 
&\quad + V_\text{S}(R) \cdot \hat{P}_\text{S},
\label{eq:1}
\end{array}
\end{equation}
where $\hat{H}_0 $ is the field-free Hamiltonian of the Rydberg electron in which the fine structure is included via quantum defects \cite{Mack2011,Li2003}, $\hat{H}_\text{B} = - (\mu_B /\hbar)\,\vec{B}\cdot(\hat{L}_\text{r} + 2 \hat{S}_\text{r}+ 2\hat{S}_\text{g})$ accounts for the interaction of the Rydberg electron orbital angular momentum and spin, as well as the ground state valence electron spin with the magnetic field and $\hat{H}_\text{HF,g} $ and $\hat{H}_\text{HF,r}$ denote the hyperfine interaction of the ground and the Rydberg state, respectively. The  last two terms account for the singlet ($S$) and triplet ($T$) scattering between Rydberg electron and ground state atom, where $V_T$, $V_S$ are the scattering potentials and  $\hat{P}_\text{T}$, $\hat{P}_\text{S}$ are the projection operators, $\hat{P}_\text{T} = \hat{S}_\text{r} \otimes \hat{S}_\text{g} + 3/4 \cdot \hat{\openone}$ and $\hat{P}_\text{S} = \hat{\openone} - \hat{P}_\text{T}$. $\hat{S}_\text{g}$ and $\hat{S}_\text{r}$ are electron spin operators of the ground state atom and the Rydberg atom, respectively. We use the basis
$\left\lbrace  \left| m_{I_\text{g}}, m_{S_\text{g}}, m_{S_\text{r}}, m_{I_\text{r}} \right> \right\rbrace$, where $m_{I_\text{g}}$, $m_{I_\text{r}}$ are the nuclear spin quantum numbers and $m_{S_\text{g}}$, $m_{S_\text{r}}$ are the electron spin quantum numbers of the ground state atom and the Rydberg atom, respectively. In the subspace of a single Rydberg $nS$-state interacting with a ground state atom, the Hamiltonian has the structure shown in Fig.\,\ref{fig1}a, where the Rydberg hyperfine interaction is omitted for simplicity. The
magnetic field term enters the Hamiltonian only on the diagonal and shifts the individual levels by the Zeeman shift. The hyperfine interaction of the ground state atom $H_\text{HF,g}=A_\text{HF,g} \cdot \hat{S}_\text{g} \otimes \hat{I}_\text{g}$, where $A_\text{HF,g}$ is the hyperfine constant \cite{Bize1999}, contributes to the diagonal and off-diagonal terms of the Hamiltonian. The Rydberg electron-ground state atom scattering enters the Hamiltonian (Eq.\,\ref{eq:1}) via pure singlet $V_S$ and pure triplet $V_T$ scattering potentials on diagonal and off-diagonal positions. Eventually, the interplay of all couplings results in a mixing of the molecular states. As can be seen in Fig.\,\ref{fig1}a, the stretched two-atom states, where all considered spins are aligned, remain isolated from the other states even with all couplings included. Previous experiments using Rydberg $S$-states exclusively investigated these states, thus providing no measurement of the singlet scattering length \cite{Bendkowsky2009,Bendkowsky2010,Gaj2014}.
\\ \indent
We calculate the Rydberg electron-ground state atom scattering in two different ways.

\subsection{Effective scattering length approach}
In the first approach we assume both singlet and triplet molecular potentials to be directly proportional to the Rydberg electron probability density
$\left| \Psi (R)\right|^2$ \cite{Fermi1934,Greene2000}:
\begin{equation}
V(R)_{S,T}=2\pi a_{S,T} \left| \Psi (R)\right|^2,
\label{eq:2}
\end{equation}
where $a_{S,T}$ are the effective zero-energy $s$-wave singlet (S) or triplet (T) scattering lengths and the dependence on the relative momentum $k$ of the colliding particles as well
as the $p$-wave and higher contributions are neglected. We extract the zero-energy effective triplet scattering length from all our previous measurements of pure triplet Rydberg $S$-state molecules, which includes states with principal quantum number $n$ ranging from 34 to 71 \cite{Bendkowsky2009,Gaj2014}. The obtained value of $a_T=-15.7(1)a_0$ is adjusted to the experimental data in a least-square fit. With this fixed value used as an input, we fit the zero-energy singlet $s$-wave scattering length to the data presented in this paper.

\subsection{Full diagonalization of the scattering Hamiltonian}
In the second approach we include partial waves up to $p$-wave \cite{Omont1977}:
\begin{equation}
V_{S,T}(\vec{r},\vec{R})=2\pi A_{0}^{S,T}(k)\delta(\vec{r}-\vec{R})+6\pi A^{S,T}_{1}(k)\overleftarrow{\nabla} \delta(\vec{r}-\vec{R}) \overrightarrow{\nabla},
\label{eq:4}
\end{equation}
where $A_l(k)^{S,T}$=$-\tan\left(\delta_l(k)^{S,T}\right)/(k^{2l+1})$ are the energy-dependent singlet/triplet scattering lengths. $\delta^{S,T}_{l=0,1}$ are scattering phase shifts provided by I. I. Fabrikant \cite{Fabrikant1986}. We use a semiclassical approximation of the relative momentum $k(R)$, $k^2/2=-1/2n^{*2}+1/R$ \cite{Greene2000} and calculate the pure singlet and triplet potentials in a basis including all Rydberg states with $\Delta n=\pm 15$ around the target state. By choosing the interatomic axis as quantization axis, the $s$-wave scattering operator has non-vanishing matrix elements for states only with $m_l=0$, while for the $p$-wave scattering only $m_l=0,\pm 1$ states contribute, greatly reducing the number of basis states. We diagonalize the scattering Hamiltonian (Eq. \ref{eq:4}) as a function of the distance $R$ between the Rydberg ionic core and the perturber and obtain the pure singlet and triplet potential energy curves (PEC).

\subsection{Molecular potential energy curves}
\begin{figure}[t]
\centering
\includegraphics[scale=0.83]{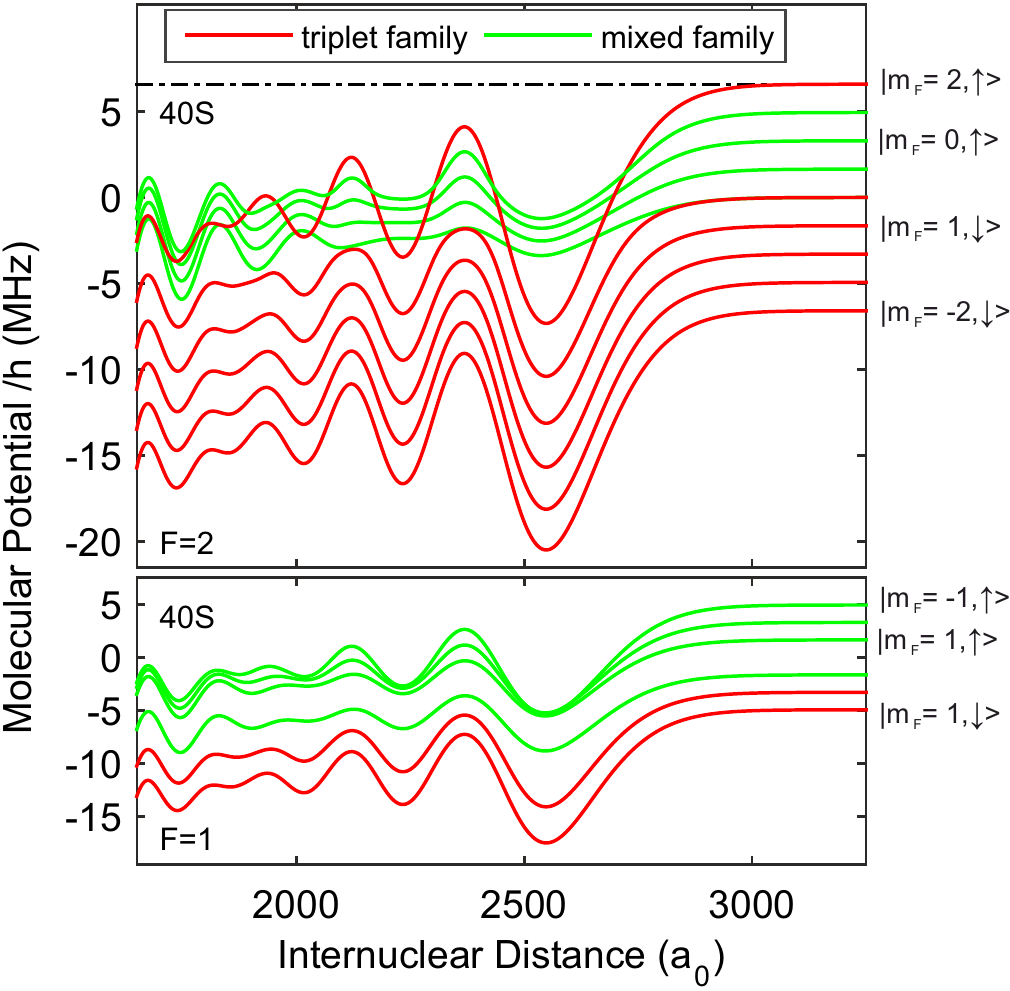}
\caption{\label{fig2} Molecular potentials for the 40$S$-5$S_{1/2}$ (F=1,\,2) molecules as a function of the internuclear distance calculated with the full diagonalization method. The inclusion of the hyperfine interaction of the ground state atom causes splitting of the molecular potential into two: triplet (red) and mixed (green) adiabatic potential energy curves. The applied magnetic field $B$=2.35\,G lifts the degeneracy of the triplet and mixed PECs and leads to further mixing of the states. Zero on the vertical axis corresponds to the asymptotes of the respective hyperfine states $F=1,2$ of the ground state in zero magnetic field. The black dashed-dotted line indicates the position of the atomic spin-up Rydberg state, which is used as reference point from now on in the paper. }
\end{figure}

Finally, we diagonalize the full Hamiltonian (Eq. \ref{eq:1}) restricted to the target state, with the singlet and triplet PEC included, at each position $R$ of the perturber. In the case of zero magnetic field we obtain two degenerate adiabatic potential energy curves: a pure triplet and a mixed singlet-triplet for both F=1 and F=2 \cite{Anderson2014b,Sassmannhausen2015}. The applied magnetic field leads to further changes of the PECs, which is shown in Fig.\,\ref{fig2} for an external magnetic field of $B$=2.35\,G. In the figure it can be seen, that the applied magnetic field lifts the degeneracy of the PECs due to the different Zeeman shift of the asymptotic atomic states, as well as the effect of further mixing of the states since the different PECs are not identical anymore.
\begin{figure}[t]
\centering
\includegraphics[scale=0.79]{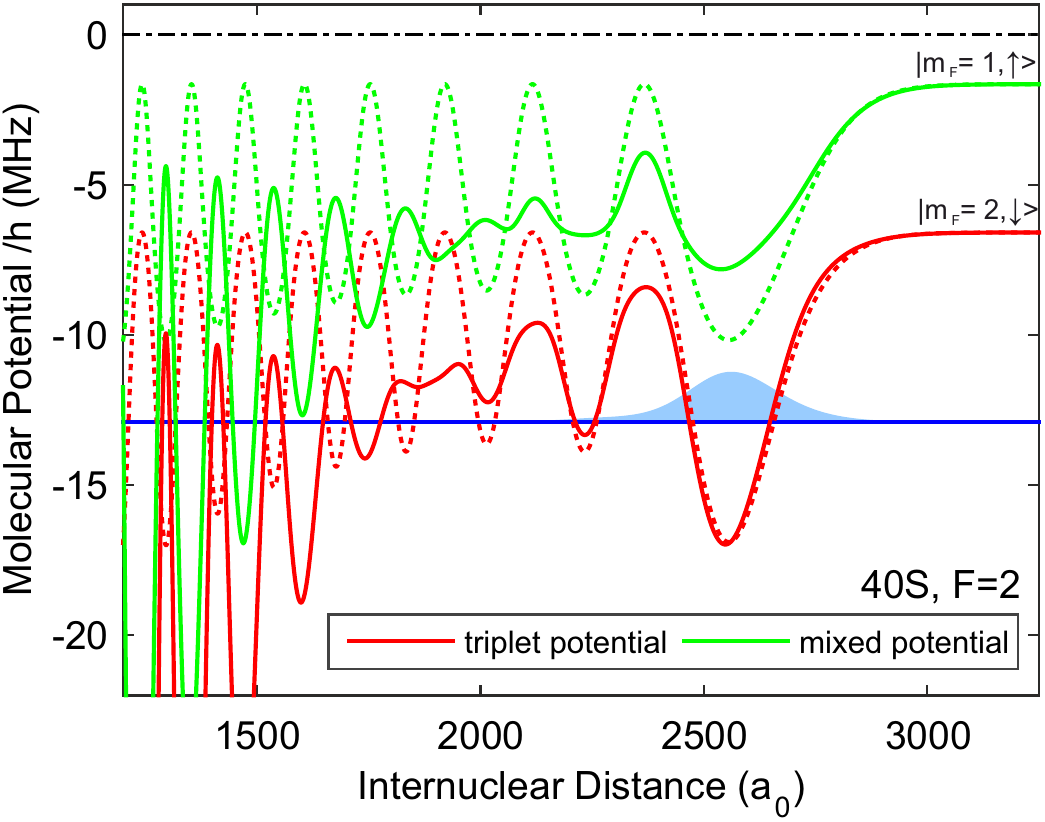}
\caption{\label{fig3} The experiment only addresses the triplet and mixed molecular states with ground state hyperfine quantum number F=2. The solid red and green lines show the respective potentials obtained with the full matrix diagonalization method, including the p-wave contribution. In contrast, the red and green dotted lines show the calculated triplet and mixed potentials using the effective scattering length approach. It can be seen that the energy dependence of the scattering length and the p-wave scattering contribution are especially important for the shallow mixed potential energy curves. Zero on the vertical axis corresponds to the atomic spin-up Rydberg state (black dashed-dotted line). As an example a bound state wavefunction inside the accessible triplet potential is plotted in blue.}
\end{figure}
This results in the case of no Rydberg hyperfine interaction in ten adiabatic potential energy curves for the 40S-5S$_{1/2}$(F=2) Rydberg molecules, which can be grouped into two families: the triplet family, where the triplet contribution to PEC is dominant and the mixed family, where the singlet contribution is significant. In the course of this paper, we keep the naming convention from the situation with zero magnetic field, hence \textit{triplet} and \textit{mixed} labels. The outermost states of F=2 have a purely triplet character and correspond to the outermost states on the diagonal of the Hamiltonian (Fig.\,\ref{fig1}a). The states of F=2, which have a nonzero overlap with the state $\left| m_{I_\text{g}}=3/2,\uparrow_g,\downarrow_r,m_{I_\text{r}}=3/2 \right\rangle$ are also shown separately in figure Fig.\,\ref{fig3} and are observed in the experiment. The $\left| m_{F_\text{g}}=1/2,\downarrow_r \right\rangle$ substate of F=1 also has a non-zero overlap with the state that is excited in the experiment. However, this overlap is very small and thus this state was not observed in the experiment. The Rydberg hyperfine interaction further splits each individual line from Fig.\,\ref{fig2} into four lines separated by around 200\,kHz, since the experiment is conducted in a magnetic field in which the Zeeman shift of the Rydberg states is large compared to their hyperfine splitting $A_{HF,r}$ \cite{Mack2011}.
\\ \indent
Subsequently, we solve the Schr\"{o}dinger equation for the nuclear motion in the calculated triplet and mixed potentials using Numerov's method to find the bound states.

\section{Molecular Spectra}
\begin{figure}[t]
\centering
\includegraphics[scale=0.87]{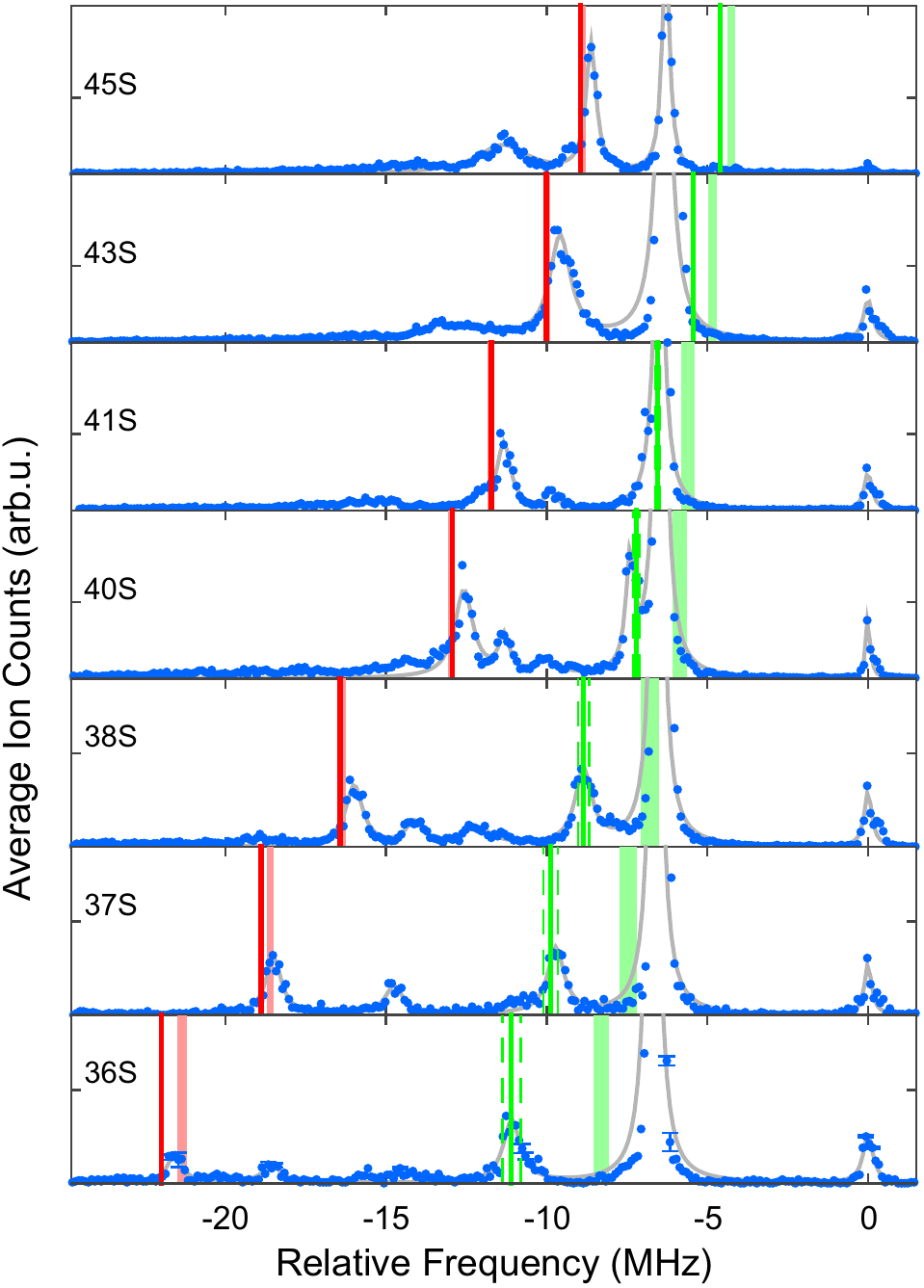}
\caption{\label{fig4} Molecular spectra for different principal quantum numbers $n$. The external applied magnetic field was $B$=2.35\,G. The zero frequency position is set to the atomic transition from the $F=2$ ground state to the $nS$ Rydberg spin-up state. The large peak at around -6.5\,MHz corresponds to the atomic transition to the spin-down Rydberg state. The red and green shaded areas show the calculated peak positions using the full diagonalization of the triplet and mixed molecular ground states, respectively. The shaded areas are used to reflect the sensitivity of the calculated binding energies to the chosen boundary conditions of the potential used in the Schr\"{o}dinger equation. Likewise, the green and red straight lines are the calculated positions of the molecular states with the effective $s$-wave scattering length approach. For the triplet scattering length we take the fixed value of 15.7(1)a$_\text{0}$, while the singlet one a$_s$=-0.2a$_0$ is fitted to the data. The green dashed line indicates the change of the singlet scattering length by $\pm$0.5\,a$_0$. The gray lines are Lorentzian fits plotted for better visibility. Each spectrum is an average of 2000-6000 measurements with the standard error (of the mean) bars shown exemplary in the spectrum of 36$S$.}
\end{figure}

In order to observe Rydberg molecules in the experiment, we prepare a spin polarized atomic sample of approximately 3$\cdot10^6$ $^{87}$Rb atoms, at a temperature below 1\,$\mu$K, in the magnetically trapped 5$S_\text{1/2}$, F=2, m$_\text{F}$=2 state. We excite the atoms with 20\,$\mu$s long pulses with a 2\,kHz repetition rate to the $nS_\text{1/2}$ Rydberg state via a two-photon transition (Fig.\,\ref{fig1}b). The laser polarization and the detuning from the intermediate state is chosen such that we mainly excite Rydberg atoms in the $\left|\downarrow_r, m_{I_\text{r}}=3/2 \right\rangle$ state. The polarization of the excitation is not perfectly set and a small fraction in the $m_{s_\text{r}}=1/2$ still remains. We use the atomic line of both spin states to calibrate the magnetic field. After each excitation pulse we field ionize the Rydberg atoms and collect the ions on a micro-channel plate detector. In a single atomic cloud we apply 1000 excitation (at a fixed laser frequency) and ionization pulses. The obtained signal is averaged over 2-6 clouds per frequency. In this way we acquire spectra presented for different principal quantum numbers $n$ in Fig.\,\ref{fig4}.
\\ \indent
Zero frequency corresponds to the atomic transition from the $F=2$ ground to the Rydberg spin-up state, for which the absolute transition frequency does not change in the magnetic field and the transition is therefore used as reference point of the measurements. The large peak appearing at around -6.5\,MHz corresponds to the atomic transition to the Rydberg spin-down state shifted by the Zeeman term in the Hamiltonian (Eq.\,\ref{eq:1}). We attribute the outermost peak at lowest frequencies, in each spectrum, to the molecular ground state bound inside the triplet potential (Fig.\,\ref{fig2}). The calculated binding energies using both effective $s$-wave scattering length and the full diagonalization method reproduce the position of the triplet molecule with a very good accuracy. The smaller peaks at higher frequencies, that are visible in the spectra for the $n$=36-41, are attributed to excited vibrational states in the triplet potential. In the spectra for $n$=41-45 peaks at even smaller frequencies than the molecular ground state are visible, that correspond to triatomic states in the triplet potential with twice the binding energy of the molecular diatomic ground state. The peak at higher frequencies visible in the spectra of 36$S$-40$S$  of similar signal strength as the triplet ground state molecule is the molecular ground state in the mixed singlet-triplet potential, which we find in the calculations to have comparable overlap with the $\left| m_{I_\text{g}}=3/2,\uparrow_g,\downarrow_r,m_{I_\text{r}}=3/2 \right\rangle$ state as the triplet molecule.
\\ \indent
We fit the zero-energy singlet $s$-wave scattering length, to match the binding energy of the mixed molecule to the experimental data, in a least square
fit obtaining the value of a$_s$=-0.2a$_0$. A change of the obtained value of the singlet $s$-wave scattering length by $\pm$0.5\,a$_0$ does not substantially influence the calculated binding energy (Fig.\,\ref{fig4}). This obtained value of the singlet scattering length is different from the value predicted for Rb in \cite{Sassmannhausen2015}, in which they did not include the interaction with the magnetic field in the theoretical model. This approach allows one to predict the position of the molecular peaks satisfactorily and can be used as an estimate for future experiments. Calculation of the potentials using the full diagonalization method reveals the importance of the $p$~-~wave scattering contribution \cite{Borodin1991,Thumm1991,Bahrim2000}. The difference in shape of the potential using the two approaches can be seen in Fig.\,\ref{fig3} and is especially pronounced for the mixed molecules. There, the $k$-dependence and the $p$-wave scattering lowers the outermost lobe of the scattering potentials, such that it becomes comparable with the second last lobe. As a consequence, the molecular ground state is delocalized over multiple wells and effectively has a lower binding energy. The full calculation, without any additional fitting parameters, reproduces the observed molecular lines with an accuracy better than $2.5$\,MHz. This shows that the nonrelativistic $e^-$\,-\,Rb scattering phase shifts must be quite accurate and are appropriate for calculating Rydberg molecule potentials. Nonetheless, the resolution achieved in our experiment reveals a systematic deviation, which suggests that our data could be used to refine these scattering phase shifts.

\section{Magnetic Field Dependency}
\begin{figure}[t]
\centering
\includegraphics[scale=0.8]{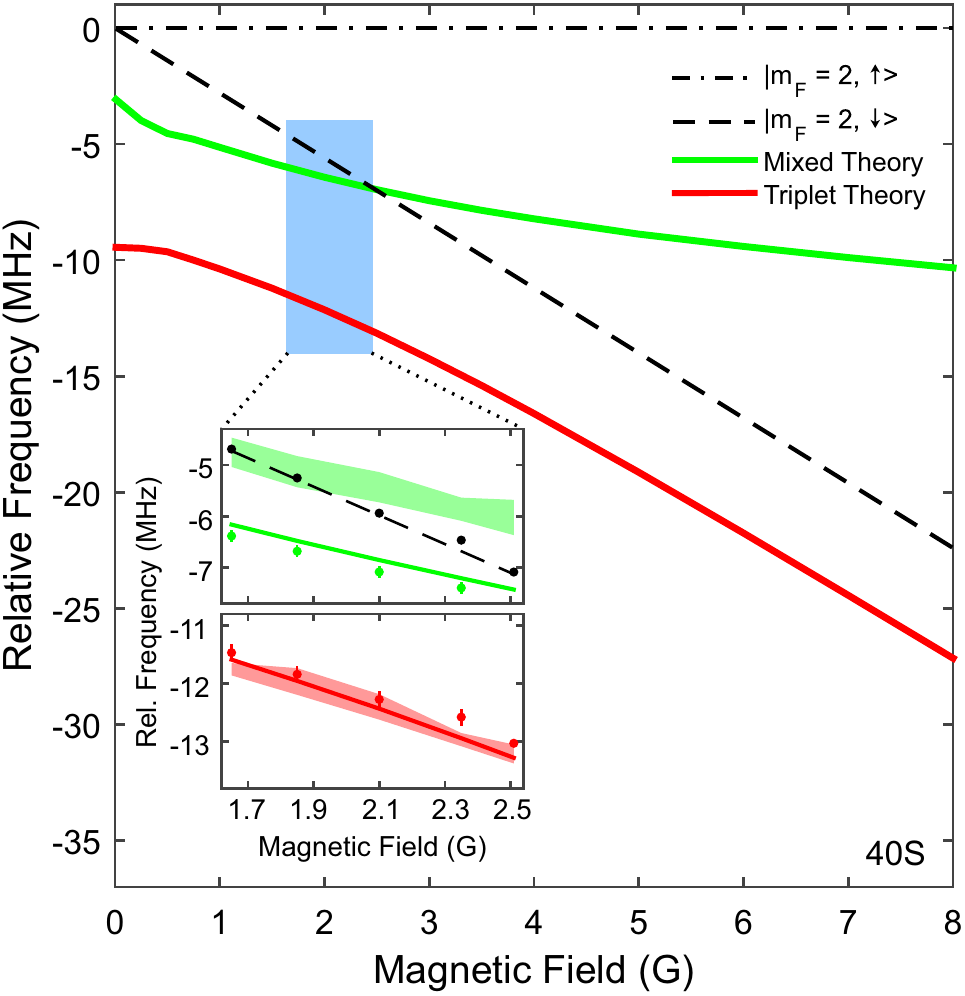} %
\caption{\label{fig5}
Theoretical Zeeman map of the mixed and triplet molecular states corresponding to the asymptotic $40$S Rydberg and $F=2$ ground state, calculated using the effective scattering length approach. Zero energy reference point is the atomic pair state $\left| m_{F_\text{g}} = 2, m_{S_\text{r}} = 1/2 \right>$. The black dashed line shows the theoretical Zeeman shift of the atomic Rydberg spin down state, which we use for the calibration of the magnetic field in the experiment. The experimental data from our spectroscopy in the magnetic trap lie in the blue shaded region and are shown as dots in the inset: atomic Rydberg spin down state (black), triplet molecular ground state (red) and mixed molecular ground state (green). Straight lines show the calculated line positions using the effective scattering length approach, shaded areas indicate the results of the full diagonalization calculation (color code is the same as for experimental data). The atomic Rydberg spin down state is again shown by the black dashed line. Error bars are the 95\% confidence intervals of the Lorentzian fit positions taken from more than 4000 measurements.}
\end{figure}

In the last section we study the magnetic field dependency of the mixed singlet-triplet molecules. Anderson et al. have used a perturbative approach to calculate magnetic moments of the zero-field pure triplet and mixed states \cite{Anderson2014b}. By including the Zeeman terms directly in the Hamiltonian (Eq.\,\ref{eq:1}) we can treat this problem non-perturbatively and calculate the full Zeeman-map for the relevant molecular states using the effective scattering length approach (Fig.\,\ref{fig5}). We find that once the Zeeman splitting becomes comparable to the energy difference between the families of pure and mixed molecular states, the shift of individual lines no longer depends linearly on the applied field. For $n$=40 this cross-over from the Zeeman to the Paschen-Back regime with respect to the molecular binding energies happens at a magnetic field $B < 1\,$G. Our experimental data falls into the "high"-field regime. We observe that the Zeeman shift of the mixed molecular states is weaker than that of two isolated atoms. This is due to all the contributions in the Hamiltonian (Eq.\,\ref{eq:1}) resulting in a mixing of the spin orientations. In the inset of Fig.\,\ref{fig5} it can be seen that the experimental data and the results of our presented theoretical model match quite well. Both of our models predict the right slope in the small range of experimentally studied values of the magnetic field, even though the calculation with the full diagonalization method shows an offset of about $1.5$\,MHz. Note that for the pure triplet states observed in previous publications, the magnetic field does not influence the shape of the molecular potential and the molecules therefore show the same Zeeman shift as the respective asymptotic atomic state.

\section{Conclusion}
In conclusion, we have investigated mixed singlet-triplet Rydberg molecules in Rb and have extracted a zero-energy singlet scattering length a$_s$=-0.2a$_0$. Inclusion of the $p$-wave term in the electron-ground state scattering reveals the importance of the $p$-wave shape resonance in particular for these mixed molecular states \cite{Bendkowsky2010}. We have studied the magnetic field dependence of the observed molecular states. The interaction with the magnetic field, similar to the electric field \cite{Li2003,Booth2015,Gaj2015}, also creates the possibility of engineering the molecular state. Our high resolution data can serve as a reference for $e^-$\,-\,Rb scattering phase shift calculations. The precision of the input parameters entering these calculations e.\,g. ground state polarizability $\alpha$, have improved over the years \cite{Gregoire2015}, which could lead to a significant change in the phase shifts we can now measure. Furthermore, the theoretical calculation can be improved further, by treating the scattering interaction and the interaction with the external magnetic field in the same step and not limiting the interaction with the magnetic field to a small subspace of the Hilbert space. In the future this can than be used to describe the magnetic field dependence of mixed singlet-triplet Rydberg molecules in the high density of a Bose-Einstein condensate, where the effect of the p-wave resonance is even more pronounced \cite{Schlagmüller2015} and as such more states have to be considered in the calculation.
																																				
\begin{acknowledgments}
We thank I. I. Fabrikant for providing us the $e^-$-Rb scattering phase shifts and Georg Raithel for very helpful comments, as well as Chris Greene and J.~P\'{e}rez-R\'{\i}os for fruitful discussions and the contribution to the full diagonalization code. We acknowledge support from Deutsche Forschungsgemeinschaft (DFG) within the SFB/TRR21 and the project PF 381/13-1. Parts of this work were also funded by ERC under contract number 267100. S.H. acknowledges support from DFG through the project HO 4787/1-1.
\end{acknowledgments}


\end{document}